\documentclass[aip,apl,reprint,longbibliography,amsmath,amssymb,floatfix,groupedaddress]{revtex4-2}

\usepackage{graphicx}
\usepackage{hyperref}

\renewcommand{\vec}{\mathbf}

\begin{document}

\title{Effective electrical conductivity of random resistor networks generated using a Poisson--Voronoi tessellation}

\author{Yuri~Yu.~Tarasevich}
\email[Corresponding author: ]{tarasevich@asu.edu.ru}
\affiliation{Laboratory of Mathematical Modeling, Astrakhan State University, Astrakhan, 414056, Russia}

\author{Irina~V.~Vodolazskaya}
\email{vodolazskaya\_agu@mail.ru}
\affiliation{Laboratory of Mathematical Modeling, Astrakhan State University, Astrakhan, 414056, Russia}

\author{Andrei~V.~Eserkepov}
\email{dantealigjery49@gmail.com}
\affiliation{Laboratory of Mathematical Modeling, Astrakhan State University, Astrakhan, 414056, Russia}

\date{\today}

\begin{abstract}
We studied the effective electrical conductivity of dense random resistor networks (RRNs) produced using a Voronoi tessellation when its seeds are generated by means of a homogeneous Poisson point process in the two-dimensional Euclidean space. Such RRNs are isotropic and in average homogeneous, however, local fluctuations of the number of edges per unit area are inevitably. These RRNs may mimic, e.g., crack-template-based transparent conductive films. The RRNs were treated within a mean-field approach (MFA). We found an analytical dependency of the effective electrical conductivity on the number of conductive edges (resistors) per unit area, $n_\text{E}$. The effective electrical conductivity is proportional to $\sqrt{n_\text{E}}$ when $n_\text{E} \gg 1$.
\end{abstract}

\maketitle

\begin{figure}[!hbt]
  \centering
  \includegraphics[width=0.8\columnwidth]{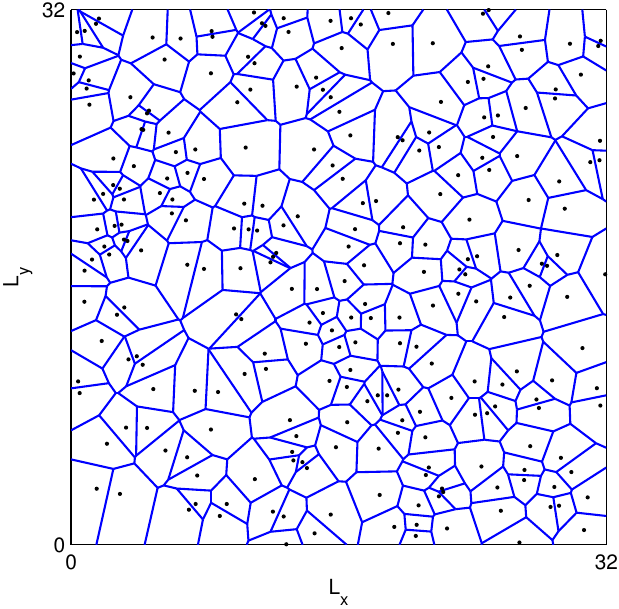}
  \caption{Example of the Poisson--Voronoi tessellation of a square domain ($L_x=L_y=32$) using 256 randomly distributed seeds (shown as points).\label{fig:PVT}}
\end{figure}

Crack-template-based transparent conductive films~\cite{Muzzillo2020b,Liu2022,Voronin20231SI} are promising kinds of junction-free, metallic network electrodes.
Recently, physical properties of crack-template-based transparent conductive films were described, while a Voronoi tessellation was applied to generate geometry of the crack patterns~\cite{Zeng2020,Kim2022,Esteki2023}. Let $S$ be a set of points on a plane. When each point of the plane is associated with the nearest point of $S$, then the space is divided into convex polygons (cells). Such a partition is known as a Voronoi (Dirichlet, Theissen) tessellation. When $S$ is generated randomly, the partition is called a Poisson--Voronoi tessellation. Figure~\ref{fig:PVT} presents an example of the Poisson-Voronoi tessellation. A local inhomogeneity can be clearly seen.

\citet{Meijering1953} reported the main statistical properties of Voronoi tessellations. It should be noted that the article~\cite{Meijering1953} used a description in terms of crystal growth rather than Voronoi diagrams. A summary can be found in~\citet{Brakke2005}, viz., each vertex has valence 3, the mean number of sides of a cell is 6, the mean cell perimeter is $4c$, whence the mean edge length is $2/3c$, where $c^2$ is the mean cell area. These properties have been confirmed using  a computer simulation of 200\,000\,000 random Voronoi polygons in the plane~\cite{Brakke2005a}. Thus, the mean edge length depends on the number density of edges as follows
\begin{equation}\label{eq:Voronoilmean}
\langle l \rangle = \frac{2}{\sqrt{3n_\text{E}}}.
\end{equation}

\citet{Kumar2016}, evaluating isotropic  RRNs from geometrical considerations, proposed an analytical formula for the sheet resistance
\begin{equation}\label{eq:KumarRs}
R_\Box = \frac{\pi}{2} \frac{\rho}{w t \sqrt{n_\text{E}}},
\end{equation}
where $\rho$ is the electric resistivity of wires, $w$ and $t$ are their width and thickness, respectively. $n_\text{E}$ is the number density of wires, i.e., the number of wires per unit area. Hence, the effective  conductivity of the RRN is
\begin{equation}\label{eq:KumarG}
\sigma_\text{eff} = \frac{2}{\pi} \sigma_0 w t \sqrt{n_\text{E}}, \quad \sigma_0 = \rho^{-1}.
\end{equation}

A similar approach has been applied to anisotropic systems~\cite{Tarasevich2019}. In this case, the sheet resistance can be written as
\begin{equation}\label{eq:Ranisotropic}
  R_\Box =\frac{2 \rho}{ w t n_\text{E} \langle l \rangle (1 \pm s)},
\end{equation}
where $l$ is the length of the crack segment.
\begin{equation}\label{eq:s}
s = 2 \langle \cos^2 \theta \rangle - 1
\end{equation}
is the orientational order parameter~\cite{Frenkel1985}, where $\theta$ is the angle between a wire and the $x$-axis. For isotropic RRNs ($s=0$), the effective electrical conductivity is
\begin{equation}\label{eq:Ganisotropic}
  \sigma_\text{eff}  =\frac{\sigma_0 w t n_\text{E} \langle l \rangle }{2 }.
\end{equation}
Equations \eqref{eq:KumarG} and \eqref{eq:Ganisotropic} differ since different assumptions regarding the number of wires intersecting a line were used.

Those approaches~\cite{Kumar2016,Tarasevich2019} leave out of account that both the electric field and the local electrical conductivity fluctuate. Indeed, since the average conductive edge length obviously depends on the edge concentration, local fluctuations in the edge concentration lead to local fluctuations in the average edge length.

\citet{OCallaghan2016} showed that the dependence of the electrical resistance on the distance between two nodes of dense RRNs behaves similarly to the dependence of resistance on the distance between two points on a conducting plane. This behavior allows us to conclude (although this is not explicitly claimed in article~\cite{OCallaghan2016}) that the continuum approximation may be appropriate to calculate the effective electrical conductivity of dense RRNs.

\citet{LL8} performed an analytical consideration of the electrical permittivity of inhomogeneous media. Very similar approach has been applied to the effective electrical conductivity of particular inhomogeneous planes~\cite{Khalatnikov2000,Barash2015}. Omitting technical detail, since detailed derivation can be found elsewhere~\cite{Khalatnikov2000,Barash2015}, the effective electrical conductivity, $\sigma_\text{eff}$, of  inhomogeneous media can be represented as follows
\begin{equation}\label{eq:sigmaeffKhalatnikov}
  \sigma_\text{eff} \vec{E} =\langle \sigma \rangle \vec{E} - \langle \sigma  \nabla \psi \rangle.
\end{equation}
Here, $\langle \cdot \rangle$ denotes a mean value, $\vec{E}$ is the mean electric field, $\sigma$ is the local, i.e., coordinate-dependent, electrical conductivity; $\sigma = \sigma(x,y)$ in 2d case.
The electric potential and the  mean electric field are connected as follows
\begin{equation}\label{eq:E}
  \varphi = - \vec{E}\vec{r} + \psi,
\end{equation}
where
\begin{equation}\label{eq:E2}
\vec{E} = - \langle \nabla \varphi \rangle,
\end{equation}
while the fluctuation $\psi = \psi(\vec{r})$ is such that its average value is zero
\begin{equation}\label{eq:psi}
\langle \nabla \psi \rangle = 0.
\end{equation}
In other words, $\vec{E}$ is the mean electric field omitting the influence of fluctuations of the electrical conductivity, i.e., when $\sigma(x,y) = \langle \sigma \rangle$.

\citet{Dykhne1971} considered the electrical conductivity of a two-dimensional two-phase system, in particular, when there is a smooth dependence of the conductivity on the coordinates. In this particular case, the effective electrical conductivity can be found as follows
\begin{equation}\label{eq:Dykhne-cont}
\sigma_\text{eff} = \exp \langle \ln \sigma \rangle = \sqrt{\langle \sigma \rangle \langle \sigma^{-1} \rangle^{-1}}.
\end{equation}

The goal of the present letter is an investigation of the electrical properties of RRNs generated using a Poisson-Voronoi tessellation.

Let there be an isotropic and in average homogeneous random conductive network of size $ L_x \times L_y$. Let a potential difference $U_0$ be applied to the opposite borders of this network along the $x$ direction. The typical edge length is assumed to be much less as compared to the linear network size, $L_x$, hence, the potential drop along the network is expected to be almost linear. Since the number density of edges is coordinate-dependent, the average number density of edges is equal to
\begin{equation}\label{eq:mean-nE}
  \langle n_\text{E} \rangle = \frac{1}{L_x L_y} \int\limits_{0}^{L_x} \int\limits_{0}^{L_y} n_\text{E}(x,y) \, \mathrm{d}x \, \mathrm{d}y = \frac{N_\text{E}}{A},
\end{equation}
where $N_\text{E}$ is the total number of edges in the network, $A = L_x L_y$ is the area of the network.
Let the electric field within the network be a normally distributed random variable with expectation
\begin{equation}\label{eq:Uloc}
  \mathbb{E} [U] = U_0\frac{x}{L_x}
\end{equation}
and variance $s^2$; the variance depends on the number density of edges. In this case, the probability density function (PDF) equals to
\begin{equation}\label{eq:PDF}
f(U) = \frac{1}{ s\sqrt{2\pi}} \exp \left[-\frac{1}{2}\left(\frac{ U - U_0 x L_x^{-1}}{s} \right)^2 \right].
\end{equation}
Let us find, as a first approximation, the effective electrical conductivity of the system, taking into account only the average electric field.

The electric current in an edge of length $l$ oriented at an angle $\alpha$ to the external electric field is equal to
\begin{equation}\label{eq:i}
i(\alpha) = \frac{U_0 l\cos \alpha}{L_x R} = \frac{U_0 w t \cos \alpha}{L_x \rho }.
\end{equation}
Here,
\begin{equation}\label{eq:R}
  R = \rho \frac{l}{w t}
\end{equation}
is the resistance of the edge, where $w$ and $t$ are the edge width and height, respectively.

Consider a line perpendicular to the external electric field. An edge of length $l$, oriented relative to the external electric field at an angle $\alpha$, intersects this line if its origin is located at a distance not exceeding $l \cos\alpha$  from the line. The segment $[y; y +\mathrm{d}y]$ intersects $n_\text{E}(x,y) l \cos \alpha \,\mathrm{d}y$ edges, while the total electric current in all the such edges is
\begin{equation}\label{eq:tmp}
\frac{U_0 n_\text{E}(x,y)  l(n_\text{E}) w t \cos^2 \alpha}{L_x \rho } \, \mathrm{d}y.
\end{equation}
$\mathrm{d}y$ is supposed to be much larger than a typical cell size, i.e., a segment of length $\mathrm{d}y$ intersects a great deal of edges.
The total electric current through the segment  $[y; y +\mathrm{d}y]$ equals to
\begin{multline}\label{eq:Itotalsimple}
\mathrm{d} I(x,y) = \frac{\mathrm{d}y}{\pi} \frac{ w t U_0}{\rho L_x } \times \\ \int\limits_{l_\text{min}}^{l_\text{max}}\int\limits_{-\pi/2}^{\pi/2}
n_\text{E}(x,y) l(n_\text{E}) f(l;n_\text{E}) \cos^2 \alpha
\,\mathrm{d}\alpha\,\mathrm{d}l.
\end{multline}
Since all edge orientations are equiprobable, while $\alpha \in [-\pi/2,\pi/2]$, the corresponding PDF is $\pi^{-1}$. $f(l;n_\text{E})$ is the PDF of the edge length, when the number density of edges is $n_\text{E}$; its support is  $l \in [l_\text{min}, l_\text{max}]$. 
Then the first approximation of the local (coordinate-dependent) electrical conductivity is
\begin{equation}\label{eq:Gsheet}
\sigma(x,y) = \frac{\sigma_0 n_\text{E}(x,y)  \left\langle l(n_\text{E}) \right\rangle w t}{2}.
\end{equation}
Thus, the first approximation of the effective electrical conductivity of the network is
\begin{equation}\label{eq:Gsheeteffcommon}
\sigma_\text{eff}^{(1)}  = \frac{\sigma_0 w t \left\langle n_\text{E}  \left\langle l(n_\text{E}) \right\rangle \right\rangle}{2}.
\end{equation}
Note that the effective electrical conductivity of the network~\eqref{eq:Gsheeteffcommon} differs from both~\eqref{eq:KumarG} and~\eqref{eq:Ganisotropic}, which were obtained without taking into account the dependence of the electric field and the local electrical conductivity of the network on coordinates.

When a network is produced using a Poisson--Voronoi tessellation, the effective electrical conductivity, accounting for~\eqref{eq:Voronoilmean}, is
\begin{equation}\label{eq:sigmaeffVoronoi0}
\sigma_\text{eff}^{(1)}  = \frac{\sigma_0 w t\langle \sqrt{n_\text{E}}\rangle}{\sqrt{3}}.
\end{equation}

Since, for large values of $\lambda$, the Poisson distribution $f(k;\lambda)$ tends to the normal distribution with mean $\lambda$ and variance $\lambda$, we suppose that the quantities of interest obey the normal distribution. Assuming that a quantity $z$ obeys the normal distribution with the expectation $\lambda$ and the variance $\lambda$, a mean value of $\sqrt{z}$ is
\begin{multline}\label{eq:meanWeber}
\langle \sqrt{z} \rangle = \frac{1}{\sqrt{2 \pi \lambda}}\int\limits_{0}^{+\infty} \sqrt{z} \exp \left(- \frac{\left(z-\lambda\right)^2}{2\lambda} \right) \, \mathrm{d}z
\\=
\frac{\lambda^{\frac{1}{4}}}{2\sqrt{2 }}
\exp \left( -\frac{\lambda}{4} \right) D_{-\frac{3}{2}}\left(-\sqrt{\lambda}\right)
\\=
\frac{\lambda}{4}
\sqrt{\frac{\pi}{2}}
\exp\left(-\frac{\lambda}{4}\right)\\
\times
\left[ I_{-\frac{3}{4}}\left(\frac{\lambda}{4} \right) + I_{\frac{3}{4}}\left(\frac{\lambda}{4} \right) +  I_{-\frac{1}{4}}\left(\frac{\lambda}{4} \right) + I_{\frac{1}{4}}\left(\frac{\lambda}{4} \right)\right].
\end{multline}
Here, $D_a(z)$ is the parabolic cylinder function (Weber function) and $I_a(z)$ is the modified Bessel function.
Hence, the effective electrical conductivity of the network is
\begin{multline}\label{eq:sigmaeffVoronoi}
\sigma_\text{eff}^{(1)}  = \frac{\sigma_0 w t}{\sqrt{3}}
\exp\left(-\frac{n_\text{E}}{4}\right)
\frac{n_\text{E}}{4} \sqrt{\frac{\pi}{2}}
\\
\times
\left[ I_{-\frac{3}{4}}\left(\frac{n_\text{E}}{4} \right) + I_{\frac{3}{4}}\left(\frac{n_\text{E}}{4} \right) +  I_{-\frac{1}{4}}\left(\frac{n_\text{E}}{4} \right) + I_{\frac{1}{4}}\left(\frac{n_\text{E}}{4} \right)\right].
\end{multline}

Accounting for~\eqref{eq:Dykhne-cont} and~\eqref{eq:sigmaeffVoronoi0}, the second approximation can be found
\begin{equation}\label{eq:sigmaeff2Voronoi}
\sigma_\text{eff}^{(2)}  = \frac{\sigma_0 w t}{\sqrt{3}} \sqrt{\langle \sqrt{n_\text{E}}\rangle \left\langle \frac{1}{\sqrt{n_\text{E}}}\right\rangle^{-1}}.
\end{equation}
Since
\begin{multline*}
\left\langle z^{-\frac{1}{2}} \right\rangle = \frac{1}{\sqrt{2 \pi \lambda}}\int\limits_0^\infty \frac{\exp \left(- \frac{\left(z-\lambda\right)^2}{2\lambda} \right)}{\sqrt{z}} \, \mathrm{d}z
\\=
\sqrt{\frac{\pi}{8}}\exp\left(-\frac{\lambda}{4} \right) \left[I_{-\frac{1}{4}}\left(\frac{\lambda}{4}\right) + I_{\frac{1}{4}}\left( \frac{\lambda}{4}\right)\right],
\end{multline*}
the next approximation of the effective electrical conductivity is
\begin{multline}\label{eq:sigmaeff2Voronoifinal}
\sigma^{(2)}_\text{eff} =\sigma_0 w t \\
\times
\left\{
\frac{n_\text{E} }{6}
\left[ 1+
\frac{
I_{-\frac{3}{4}}\left(\frac{n_\text{E}}{4} \right) + I_{\frac{3}{4}}\left(\frac{n_\text{E}}{4} \right)
}
{
I_{-\frac{1}{4}}\left(\frac{n_\text{E}}{4}\right) + I_{\frac{1}{4}}\left( \frac{n_\text{E}}{4}\right)
}
\right]\right\}^{1/2}.
\end{multline}
So far as
\begin{equation}\label{eq:limit}
\lim_{n_\text{E} \to \infty} \frac{
I_{-\frac{3}{4}}\left(\frac{n_\text{E}}{4} \right) + I_{\frac{3}{4}}\left(\frac{n_\text{E}}{4} \right)
}
{
I_{-\frac{1}{4}}\left(\frac{n_\text{E}}{4}\right) + I_{\frac{1}{4}}\left( \frac{n_\text{E}}{4}\right)} =1,
\end{equation}
when the value of $n_\text{E}$ increases, the effective electrical conductivity tends to the value predicted by the simplest version of the MFA~\cite{Tarasevich2019}, i.e., \eqref{eq:Ganisotropic} along with~\eqref{eq:Voronoilmean}
\begin{equation}\label{eq:GVoronoilimit}
\sigma^{(2)}_\text{eff} =\sigma_0 w t \sqrt{ \frac{n_\text{E}}{ 3 }}.
\end{equation}

To check the formulas for the effective electrical conductivity, we performed a comparison with direct computations. To generate networks, we used a domain of size $l_x = L_y =32$. Seeds were randomly placed within this domain. For each decided number of seeds, 10 different networks were generated. Each edge was considered as a conductive wire of the width $w$ and thickness $t$, which material owning the electrical conductivity $\sigma_0$. To reveal a dependence of the effective electrical conductivity on the number density of edges, we used a reduced conductivity, vis.,
\begin{equation}\label{eq:reducedG}
  \frac{\sigma_\text{eff}}{\sigma_0 w t}.
\end{equation}
This reduced conductivity was averaged over 10 realizations for each number density of edges.

\begin{figure}[!htb]
  \centering
  \includegraphics[width=\columnwidth]{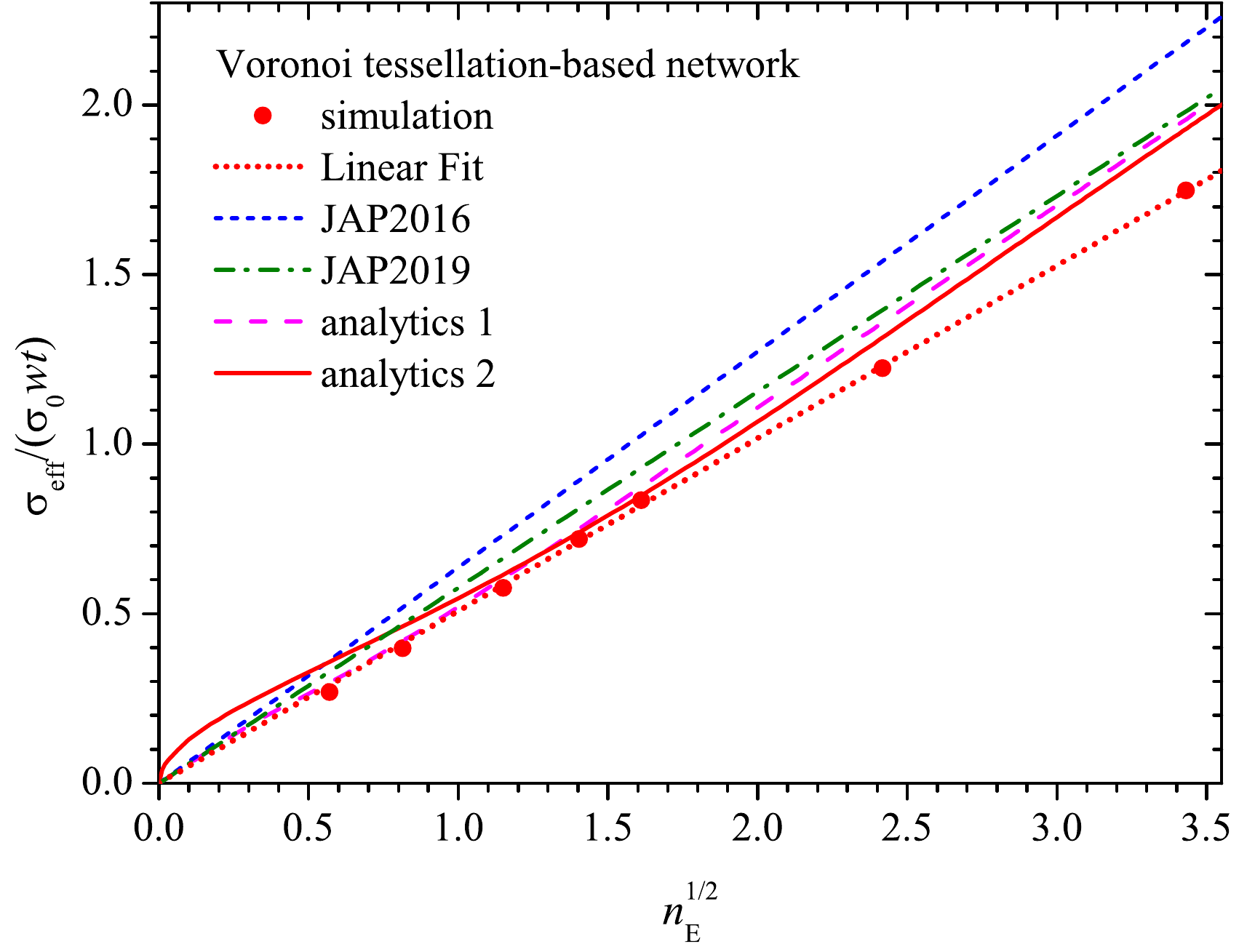}
  \caption{Reduced effective electrical conductivity of the RRNs vs the number density of resistors. Markers correspond to direct computations averaged over 10 different RRNs, while lines correspond to different approximations. The standard deviation of the mean is of the order of the marker size.\label{fig:Voronoi}}
\end{figure}
Figure~\ref{fig:Voronoi} compares predictions with the direct computations of the electrical conductivity. JAP2016 corresponds to \citet{Kumar2016}, Eq.~\eqref{eq:KumarG} in this work (slope is $2/\pi \approx 0.6366$), JAP2019 corresponds to \citet{Tarasevich2019}, Eq.~\eqref{eq:GVoronoilimit} (slope is $3^{-1/2} \approx 0.57735$), analytics~1 corresponds to~\eqref{eq:sigmaeffVoronoi}, and analytics~2 corresponds~\eqref{eq:sigmaeff2Voronoifinal}. Accounting for spatial fluctuations both the electric field and the conductivity have very small impact on the effective electrical conductivity. Thus, the standard deviations between seven computed values for computer-generated networks and analytical formulas \eqref{eq:sigmaeffVoronoi} and \eqref{eq:sigmaeff2Voronoifinal} are $0.0396245$ and $0.0360748$, respectively. Moreover, when $n_\text{E} \gtrapprox 10$, a simplest version of the MFA~\eqref{eq:GVoronoilimit} is almost as accurate as the most rigorous consideration~\eqref{eq:sigmaeff2Voronoifinal}. Linear least squares fit of the direct computations has the slope $ 0.5087  \pm 0.0027$, while the coefficient of determination is $R^2 =  0.9998$.

By means of a mean-field approach, we studied the effective electrical conductivity of dense 2d RRNs produced using a Poisson--Voronoi tessellation. Although such RRNs are isotropic and in average homogeneous, local fluctuations of the number of edges per unit area are inevitably. These RRNs may mimic, e.g., crack-template-based transparent conductive films. We found an analytical dependency of the effective electrical conductivity on the number of conductive edges per unit area, $n_\text{E}$. The effective electrical conductivity is proportional to $\sqrt{n_\text{E}}$ when  $n_\text{E} \gg 1$. We compared our formula with other formulas proposed in the literature as well with the direct computations of the  effective electrical conductivity of dense RRNs. The comparison evidenced that our formula is more accurate, however all theoretical descriptions overestimate the effective electrical conductivity even for very dense RRNs, where the MFA is expected to be rather accurate. We suggest that this deviation may be due to local anisotropy of RRNs. 

The data that support the findings of this study are available from the corresponding author upon reasonable request. 

We acknowledge funding from the Russian Science Foundation, Grant No.~23-21-00074.

\bibliography{perturbationR0}

\end{document}